\documentclass[12pt]{article}
\usepackage{amsmath}
\usepackage[mathcal]{euscript}
\usepackage{graphicx}
\def \myfigures #1#2#3#4#5#6#7#8
{\begin{figure}[ht]
    \begin{center}
        \includegraphics[width=#2 \textwidth]{#1.eps}
        \hfill
        \includegraphics[width=#6 \textwidth]{#5.eps}
        \parbox[t]{#4\textwidth}{\caption {#3}\label{#1}}
        \hfill
        \parbox[t]{#8\textwidth}{\caption {#7}\label{#5}}
    \end{center}
\end{figure} }

\newcommand {\bp}{\bar \psi}

\begin{document}
\title{Bianchi type-I model with cosmic string in the presence of a
magnetic field: spinor description}

\author{Bijan Saha
\thanks{E-mail:~~~bijan@jinr.ru;
URL: http://wwwinfo.jinr.ru/\~\,\!bijan/}\\
{\small \it Laboratory of Information Technologies}\\
{\small \it Joint Institute for Nuclear Research, Dubna}\\
{\small \it 141980 Dubna, Moscow region, Russia}
\and
Mihai Visinescu
\thanks{E-mail:~~~mvisin@theor1.theory.nipne.ro;
URL: http://www.theory.nipne.ro/\~\,\!mvisin/}\\
{\small \it Department of Theoretical Physics}\\
{\small \it National Institute for Physics and Nuclear Engineering}\\
{\small \it Magurele, P. O. Box MG-6, RO-077125 Bucharest, Romania}
}
\date{}

\maketitle

\begin{abstract}

A Bianchi type-I  cosmological model in the presence of a magnetic
flux along a cosmic string is investigated. A nonlinear spinor
field is used to simulate the cosmological cloud of strings.
It is shown that the
spinor field simulation offer the possibility to solve the system
of Einstein's equation without any additional assumptions. It is
shown that the present model is nonsingular at the end of the evolution
and does not allow the anisotropic Universe to turn into an isotropic one.

Pacs: 95.30.Sf; 98.80.Jk; 04.20.Ha

\end{abstract}

\section{Introduction}

Though the present day Universe is well described by an isotropic
and homogeneous Friedmann-Robertson-Walker (FRW) model, there are
serious theoretical arguments about the existence of an anisotropic
phase in the evolution of the Universe \cite{misner}. These ideas
were further supported by the observational data from COBE (Cosmic
Background Explorer) and WMAP (The Wilkinson Microwave Anisotropy
Probe) where small anisotropy in the microwave background radiation
was found. These lead many cosmologists to consider the anisotropic
model to describe the initial phase of the evolution which
eventually decays into an isotropic FRW one.

After the famous paper by A. Guth \cite{guth}, scalar field was
hugely used in simulating different cosmological models. But the
question occurs, if other fields can contribute to the evolution of
the Universe. As an alternative the spinor field was used due to its
sensitivity to the gravitational one
\cite{henprd,sahaprd,greene,SBprd04,kremer1,ECAA06,sahaprd06,BVI}.
The nonlinear spinor field proved  to  be able to describe different
cosmological models. For example it was shown that a suitable choice
of nonlinearity (i) accelerates the isotropization process, (ii)
gives rise to a singularity-free Universe and (iii) generates late
time acceleration.

Moreover, it is shown that a nonlinear spinor field can be used to
simulate a perfect fluid from ekpyrotic matter to phantom matter
\cite{shikin,spinpf0,spinpf}.

At the same time the string cosmological models have been used in
attempts to describe the early Universe and to investigate
anisotropic dark energy component including a coupling between dark
energy and a perfect fluid (dark matter) \cite{KM,KM1}. Cosmic
strings are one dimensional topological defects associated with
spontaneous symmetry breaking in gauge theories. Their presence in
the early Universe can be justified in the frame of grand unified
theories (GUT).

On the other hand, the magnetic field has an important role at the
cosmological scale and is present in galactic and intergalactic spaces.
Any theoretical study of cosmological models which contain a magnetic
field must take into account that the corresponding Universes are
necessarily anisotropic. Among the anisotropic spacetimes, Bianchi
type-I space (BI) seems to be the most convenient for testing different
cosmological models.

The object of this paper is to investigate a BI string
cosmological model in the presence of a magnetic flux. For this purpose
we use the nonlinear spinor field simulation as was described recently
in a number of papers \cite{shikin,spinpf0,spinpf}.

\section{Basic equations}\label{basic}

We study the evolution of the Universe in presence of a cosmic
string and a magnetic flux in the framework of a BI
anisotropic cosmological model. For a BI spacetime
the line element is given by
\begin{equation}
ds^2 =  (dt)^2 - a_1^2(t) (dx^1)^2 - a_2^2(t)(dx^2)^2 - a_3^2(t)
(dx^3)^2\,. \label{BI}
\end{equation}
There are three scale factors $a_i$ $(i=1,2,3)$ which are functions
of time $t$ only and consequently three expansion
rates. In principle all these scale factors could be different and it
is useful to express the mean expansion rate in terms of the average
Hubble rate:
\begin{equation}\label{Hubble}
H = \frac{1}{3}\Bigl(\frac{\dot a_1}{a_1}+\frac{\dot a_2}{a_2}+
\frac{\dot a_3}{a_3}\Bigr)\,,
\end{equation}
where over-dot means differentiation with respect to $t$.

Einstein's gravitational field equations, corresponding to the
metric \eqref{BI} have the form
\begin{subequations}
\label{BID}
\begin{eqnarray}
\frac{\ddot a_2}{a_2} +\frac{\ddot a_3}{a_3} + \frac{\dot a_2}{a_2}\frac{\dot
a_3}{a_3}&=&  \kappa T_{1}^{1}\,,\label{11}\\
\frac{\ddot a_3}{a_3} +\frac{\ddot a_1}{a_1} + \frac{\dot a_3}{a_3}\frac{\dot
a_1}{a_1}&=&  \kappa T_{2}^{2}\,,\label{22}\\
\frac{\ddot a_1}{a_1} +\frac{\ddot a_2}{a_2} + \frac{\dot a_1}{a_1}\frac{\dot
a_2}{a_2}&=&  \kappa T_{3}^{3}\,,\label{33}\\
\frac{\dot a_1}{a_1}\frac{\dot a_2}{a_2} +\frac{\dot
a_2}{a_2}\frac{\dot a_3}{a_3}+\frac{\dot a_3}{a_3}\frac{\dot
a_1}{a_1}&=&  \kappa T_{0}^{0}\,, \label{00}
\end{eqnarray}
\end{subequations}
where $\kappa$ is the gravitational constant. The energy momentum tensor
for a system of cosmic string and magnetic field in a comoving
coordinate is given by
\begin{equation}
T_{\mu}^{\nu} =  \rho u_\mu u^\nu - \lambda x_\mu x^\nu
+ E_\mu^\nu\,, \label{imperfl}
\end{equation}
where $\rho$ is the rest energy density of strings with massive
particles attached to them and can be expressed as $\rho = \rho_{p} +
\lambda$, where $\rho_{p}$ is the rest energy density of the
particles attached to the strings and $\lambda$ is the tension
density of the system of strings \cite{letelier,pradhan,tade}
which may be positive or negative. Here $u_i$ is the four
velocity and $x_i$ is the direction of the string, obeying the
relations
\begin{equation}
u_iu^i = -x_ix^i = 1, \quad u_i x^i = 0\,. \label{velocity}
\end{equation}

In \eqref{imperfl} $E_{\mu\nu}$ is the electromagnetic field
\cite{lich} and in  what follows we shall choose the string and
also the magnetic field along $x^1$ direction.
In our model the electromagnetic field tensor
$F^{\alpha \beta}$ has only one non-vanishing component, namely
\begin{equation}
F_{23} = h\,,
\label{f23}
\end{equation}
where $h$ is presumed to be constant.
For the electromagnetic field $E_\mu^\nu$ one gets the following
non-trivial components
\begin{equation}
E_0^0 = E_1^1 = - E_2^2 = - E_3^3 = \frac{h^2}
{2 {\bar \mu} a_2^2 a_3^2}
\equiv \frac{1}{2}\frac{\beta^2}{(a_2 a_3)^2}\,.
\label{E}
\end{equation}
where $\bar \mu$ is a constant characteristic of the medium and called
the magnetic permeability. Typically $\bar \mu$ differs from unity
only by a few parts in $10^{-5}$ ($\bar \mu > 1$ for paramagnetic
substances and $\bar \mu < 1$ for diamagnetic). To simplify the
notation, we include in a constant $\beta$ the value of the
electromagnetic field, $h$, and the magnetic permeability, $\bar \mu$.

Using comoving coordinates we have the following components of
energy momentum tensor \cite{ass}:
\begin{equation}\label{total}
T_{0}^{0} - \rho = T_{1}^{1} - \lambda = - T_{2}^{2} =
- T_{3}^{3} =  \frac{\beta^2}{2}\frac{a_1^2}{\tau^2}\,,
\end{equation}
where we introduce the volume scale of the BI space-time \cite{sahaprd}
\begin{equation}
\tau = \sqrt{-g} = a_1 a_2 a_3 \,, \label{taudef}
\end{equation}
which is connected with the Hubble rate \eqref{Hubble}, namely
$
\frac{\dot \tau}{\tau} = 3 H\,.
$

In view of $T_{2}^{2} = T_{3}^{3}$ from \eqref{total} one
finds
\begin{equation}
a_2 = a_3 D \exp \biggl(X \int \frac{dt'}{\tau}
\biggr)\,, \label{a2a3}
\end{equation}
with $D$ and $X$ some integration constants.

On the other hand, summation of Einstein equations
\eqref{11}, \eqref{22}, \eqref{33} and three times \eqref{00} gives:
\begin{equation}
\frac{\ddot \tau}{\tau}= \frac{1}{2}\kappa \Bigl(3\rho + \lambda +
\beta ^2 \frac{a_1^2}{\tau^2} \Bigr)\,.
\label{dtau1}
\end{equation}

Taking into account the conservation of the energy-momentum tensor,
i.e., $T_{\mu;\nu}^{\nu} = 0$, after a little manipulation of
\eqref{total} one obtains \cite{SV,SRV}:
\begin{equation}\label{rholambda}
\dot \rho + \frac{\dot \tau}{\tau}\rho - \frac{\dot a_1}{a_1}\lambda
= 0\,.
\end{equation}

It is customary to
assume a relation between $\rho$ and $\lambda$ in accordance with
the state equations for strings.
The simplest one is a proportionality relation \cite{letelier}:
\begin{equation}\label{rhoalphalambda}
\rho = \alpha \lambda \,.
\end{equation}
Among the most usual choices of the constant $\alpha$  we mention
the following:
\begin{equation}\label{alpha}
\alpha =\left \{
\begin{array}{ll}
1,& \quad {\rm geometric\,\,\,string}\\
1 + \omega, & \quad \omega \ge 0, \quad p \,\,{\rm
string\,\,\,or\,\,\,
Takabayasi\,\,\,string}\\
-1, & \quad {\rm Reddy\,\,\,string}
\end{array}
\right.
\end{equation}

It is possible to consider a more general barotropic relation
\begin{equation}\label{rholambdagen}
\rho = \rho ( \lambda) \,.
\end{equation}
than the linear relation \eqref{rhoalphalambda}, subject to the
restrictions imposed by the energy conditions. The weak energy condition
as well as the strong one require $\rho \geq \lambda$ with $\lambda
\geq 0$ or $\rho \geq 0$ with $\lambda < 0$ and the dominant energy
condition implies $\rho \geq 0$ and $\rho^2 \geq \lambda^2$
\cite{letelier}.

\section{Spinor field approach}

Recently it was shown that a nonlinear spinor can be used to
simulate different types of perfect fluids including those called
ekpyrotic, phantom matter and dark energy
\cite{shikin,spinpf0,spinpf}. Here we show that it is possible to
describe cosmic strings in terms of spinor fields as well.

\subsection{Spinor field simulation}

We shall simulate the cloud formed by massive cosmic strings with
particles attached along their extensions with a nonlinear spinor field
described by the Lagrangian:
\begin{equation}
L_{\rm sp} = \frac{i}{2} \biggl[\bp \gamma^{\mu} \nabla_{\mu} \psi-
\nabla_{\mu} \bar \psi \gamma^{\mu} \psi \biggr] - m\bp \psi + F\,,
\label{lspin}
\end{equation}
with $F$ being some arbitrary function of the scalar $S = \bar \psi\psi$.
For the simulation of the present cosmological model in which
the anisotropic scale factors $a_i$ are functions
solely of time it is adequate to assume
that the spinor field depends on $t$ only. Then the
corresponding components of energy-momentum tensor take the form
\begin{subequations}
\begin{eqnarray}
T_0^0 &=& mS - F\,, \label{t00s}\\
T_i^i &=& S \frac{dF}{dS} - F\,\,\,, i=1,2,3\,.  \label{t11s}
\end{eqnarray}
\end{subequations}

In what follows we describe the energy density $\rho$ of the string by
$ T_0^0$ and the tension density $\lambda$ by $ T_1^1$ in agreement
with \eqref{total}.
Inserting \eqref{t00s} and \eqref{t11s} into \eqref{rhoalphalambda}
we find
\begin{equation}
S \frac{dF}{dS} - (1 -\frac{1}{\alpha})F - \frac{m}{\alpha} S= 0\,,
\label{eos1s}
\end{equation}
with the solution
\begin{equation}
F = C_1 S^{(\alpha - 1)/\alpha} + mS\,. \label{sol1}
\end{equation}
Here $C_1$ is an integration constant. The positivity of $T_0^0$
imposes some restriction on $C_1$, namely $C_1 \le 0$. Setting $C_1
= - \nu$ we find the cosmic string can be described by the spinor
field Lagrangian
\begin{equation}
L_{\rm sp} = \frac{i}{2} \biggl[\bar \psi \gamma^{\mu} \nabla_{\mu}
\psi- \nabla_{\mu} \bar \psi \gamma^{\mu} \psi \biggr] - \nu
S^{(\alpha - 1)/\alpha}\,, \label{lspincs}
\end{equation}
remarking the disappearance of the mass term \cite{shikin}.

With these preparatives we get
\begin{subequations}
\begin{eqnarray}
\rho = T_0^0 &=& \nu S^{(\alpha - 1)/\alpha}\,, \label{t00sf}\\
\lambda =  T_1^1 &=& \frac{\nu}{\alpha} S^{(\alpha - 1)/\alpha}\,.
\label{t11sf}
\end{eqnarray}
\end{subequations}
Variation on \eqref{lspin} with respect to $\psi$ and $\bar \psi$
gives spinor field equations with nonlinear terms. On the other hand
from the spinor field equations for $S$ one finds \cite{sahaprd}
\begin{equation}
\dot S + \frac{\dot \tau}{\tau} \, S = 0\,,\label{S}
\end{equation}
with the solution
\begin{equation}
S = \frac{C_0}{\tau}\,,
\end{equation}
$C_0$ being a constant.

Taking into account this simple behavior of $S$ we have finally
\begin{subequations}
\begin{eqnarray}
\rho &=& R \tau^{-\frac{\alpha -1}{\alpha}}\,,  \label{rho}\\
\lambda &=& \frac{R}{\alpha}\tau^{-\frac{\alpha - 1}{\alpha}}\,,
\label{lambda}
\end{eqnarray}
\end{subequations}
with the constant $ R = \nu C_0^{\frac{\alpha - 1}{\alpha}}$.

Using the above formulas for $\rho$ and $\lambda$, from
\eqref{rholambda} we can determine the anisotropic factor $a_1$:
\begin{equation}\label{a_1}
a_1 = A_1\tau\,,
\end{equation}
$A_1$ being a constant of integration.
On the other hand, from equations \eqref{taudef} and \eqref{a2a3} we obtain
\begin{equation}\label{a_2}
a_2 = \sqrt{\frac{D}{A_1}} \exp \biggl(\frac{X}{2} \int
\frac{dt'}{\tau}\biggr)\,,
\end{equation}
and
\begin{equation}\label{a_3}
a_3 = \frac{1}{\sqrt{A_1 D}} \exp \biggl(-\frac{X}{2} \int
\frac{dt'}{\tau}\biggr)\,.
\end{equation}

\subsection{Asymptotic behavior}\label{asymptotic}

In what follows, we study the equation for $\tau$ in details. Using
the above expressions for $\rho, \lambda , a_1$ we get from
\eqref{dtau1}:
\begin{equation}
\ddot \tau= \frac{\kappa}{2}\frac {3 \alpha + 1}{\alpha}R
\tau^{\frac{1}{\alpha}} + \frac{\kappa \beta^2 A_1^2 }{2}
\tau\,.\label{dtaunu}
\end{equation}
We can evaluate the derivative of $\tau$ with respect to $t$ which
leads finally to a solution in quadrature
\begin{equation}
\int \frac{d\tau}{\sqrt{[\kappa R(3\alpha+1)/(\alpha +
1)]\tau^{(\alpha + 1)/\alpha} + [\kappa \beta^2 A_1^2/2]  \tau^2 +
C}} = t + t_0\,, \label{1st}
\end{equation}
$C$ and $t_0$ being some integration constants.

In spite of the fact that this equation cannot be explicitly solved,
the asymptotic behavior of the solutions for $t\rightarrow \infty$
could be found.

\subsubsection{Case I}\label{caseI}

In most cases, \eqref{1st} provides a standing expansion of the volume
scale of the BI spacetime for $t$ growing.

Indeed, for
\begin{equation}\label{1-1}
\frac{\alpha + 1}{\alpha} \leq 2 \,,
\end{equation}
i.e. $\alpha \geq 1$ or $\alpha < 0$, the term with $\tau^2$ at the
denominator of the l. h. s. of \eqref{1st} is dominant and we get an
exponential behavior
\begin{equation}\label{tauexp}
\tau \propto \exp t \,,
\end{equation}
for large $t$. In this case, only the anisotropic factor $a_1$ presents
an exponential increase for $t \rightarrow \infty$, while the scale factors
$a_2$ and $a_3$ tend to constants.

Concerning the asymptotic behavior of $\rho$ and $\lambda$ we infer from
\eqref{rho}, \eqref{lambda} that they tend to zero as
\begin{equation}\label{rholambdainfty}
\rho\,, \lambda \propto \frac{1}{\exp \left (\frac{\alpha -1}{\alpha} t
\right ) } \,.
\end{equation}

\subsubsection{Case II}

For $0 < \alpha < 1$, there is no solution with a $\tau \rightarrow
\infty$ behavior for  $t \rightarrow \infty$. In this case from
\eqref{1st} one finds:
\begin{equation}
\tau^{\frac{\alpha -1}{2 \alpha}} \propto t. \nonumber
\end{equation}

Since $\alpha -1 < 0$, $\tau$ cannot tend to infinity as $t \to \infty$.
Therefore the model does not admit a consistent solution for
$0 < \alpha < 1$ in agreement with the discussion of the general
barotropic equation \eqref{rholambdagen} from Section \ref{basic}.

\bigskip

Taking into account the absence of consistent solutions for
$0 < \alpha < 1$ as it was shown above, for the present spinor
simulation, in what follows we shall refer only to the situations
described in subsection \ref{caseI} (Case I).
It is worth also mentioning that the present model does not support
in any case a vanishing of $\tau$ for $t \rightarrow \infty$.

\subsection{Numerical simulations}

In this subsection we graphically illustrate the evolution of energy
density $\rho$, volume scale $\tau$ and metric functions $a_1$,
$a_2$ and $a_3$. In doing so we rewrite the equation for $\tau$
\eqref{dtaunu} in terms of the Hubble parameter $H$ and introduce a
new function $T$:
\begin{subequations}
\label{system}
\begin{eqnarray}
\dot \tau &=& 3H\tau, \\
\dot H &=& - 3H^2 + \frac{\kappa}{6}\frac {3 \alpha + 1}{\alpha}R
\tau^{\frac{1 - \alpha}{\alpha}} + \frac{\kappa \beta^2 A_1^2 }{6},\\
\dot T &=& \frac{T}{\tau}.
\end{eqnarray}
\end{subequations}
We also rewrite the metric functions in terms of $T$, which now read
\begin{equation}
a_1 = A_1\tau, \quad a_2 = \sqrt{\frac{D}{A_1}} T^{X/2}, \quad a_3 =
\sqrt{\frac{1}{A_1 D}} T^{-X/2}. \label{mfnew}
\end{equation}

Let us now numerically solve the system \eqref{system}. In doing so
for simplicity we set $\kappa = 1$, $A_1 = 1$, $D = 1$ and $X =1$.
In Fig. \ref{rhosppf0} we plot the evolution of energy density
$\rho$ for different values of $\alpha$, namely for $\alpha = -1$
and $\alpha = 1.5$. Evolution of $\tau$ corresponding to these
values of parameters in shown in Fig. \ref{tausppf}. For simplicity
we also set the initial values of $\tau$, $H$ and $T$ to be unity.

\myfigures{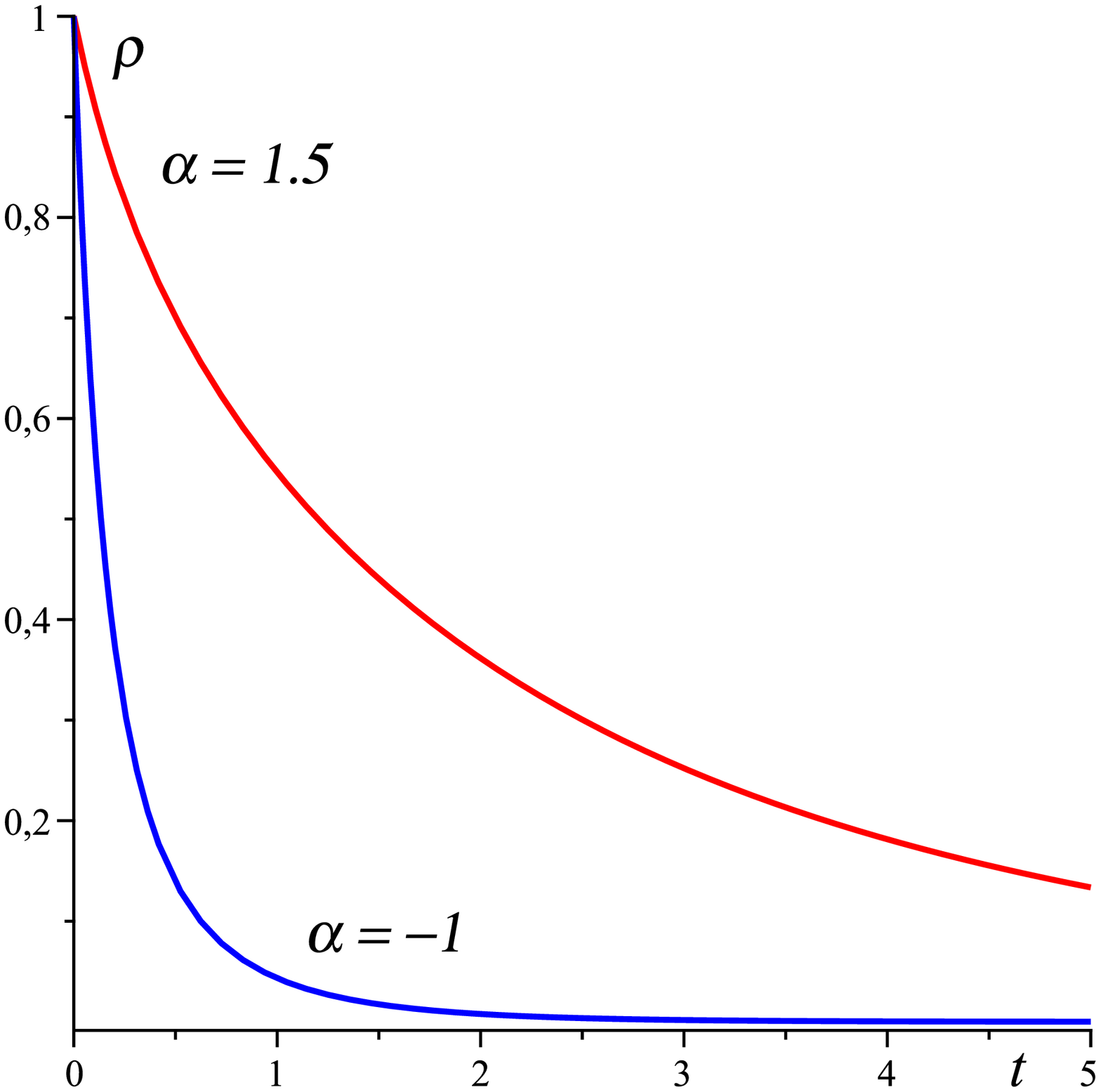}{0.45}{View of energy density of the cosmic
string for different value of $\alpha$.}
{0.45}{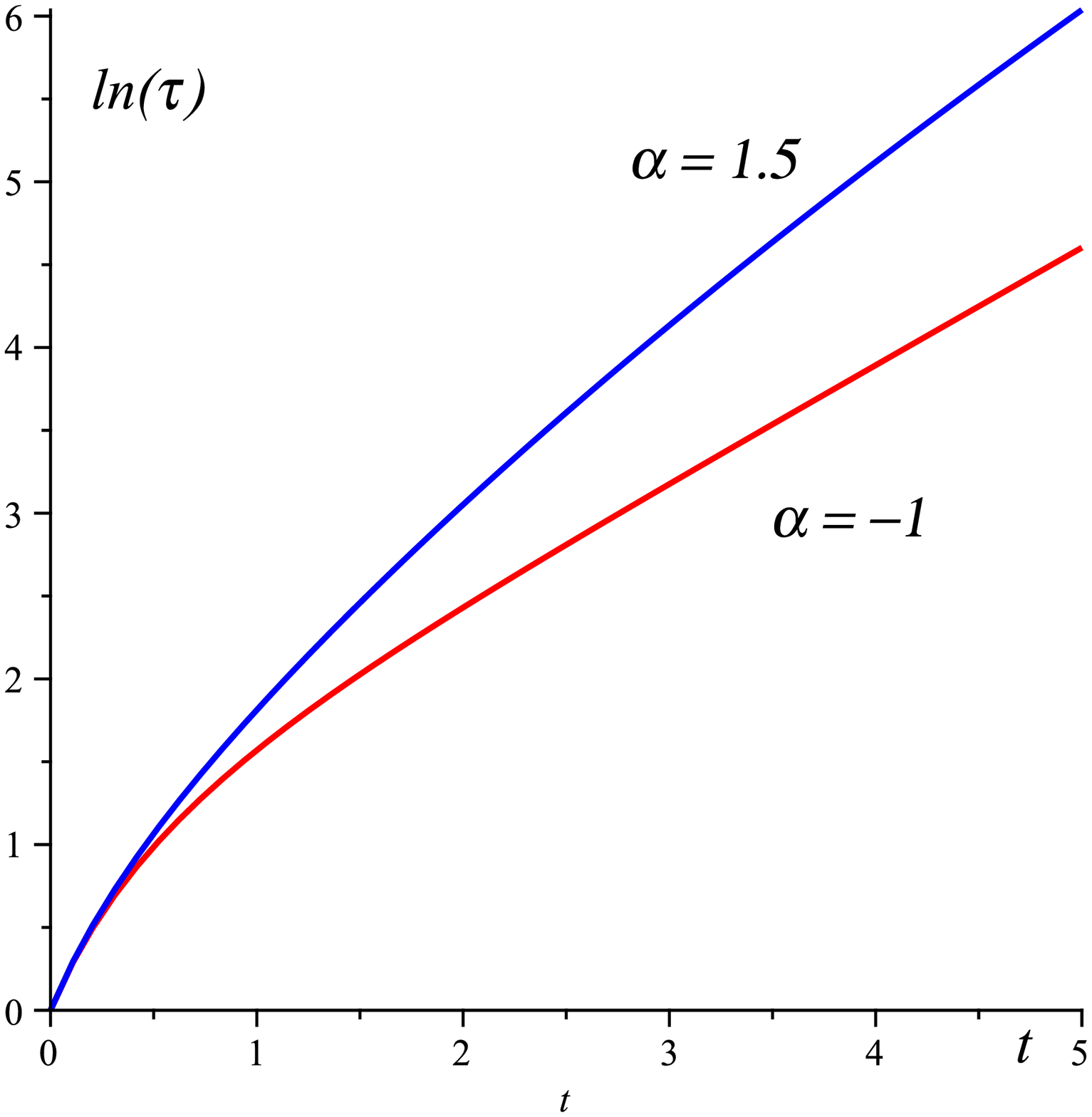}{0.45}{Evolution of the Universe corresponding to the
energy densities given in Fig. \ref{rhosppf0}.}{0.45}

The numerical simulations support the behavior described in
\eqref{tauexp}, \eqref{rholambdainfty}.
From Fig. \ref{tausppf} one finds at first, in the  case of a negative
$\alpha$ the Universe expands slower than it does as for a positive $\alpha$,
though in both cases we have an exponential growth for large time.
\myfigures{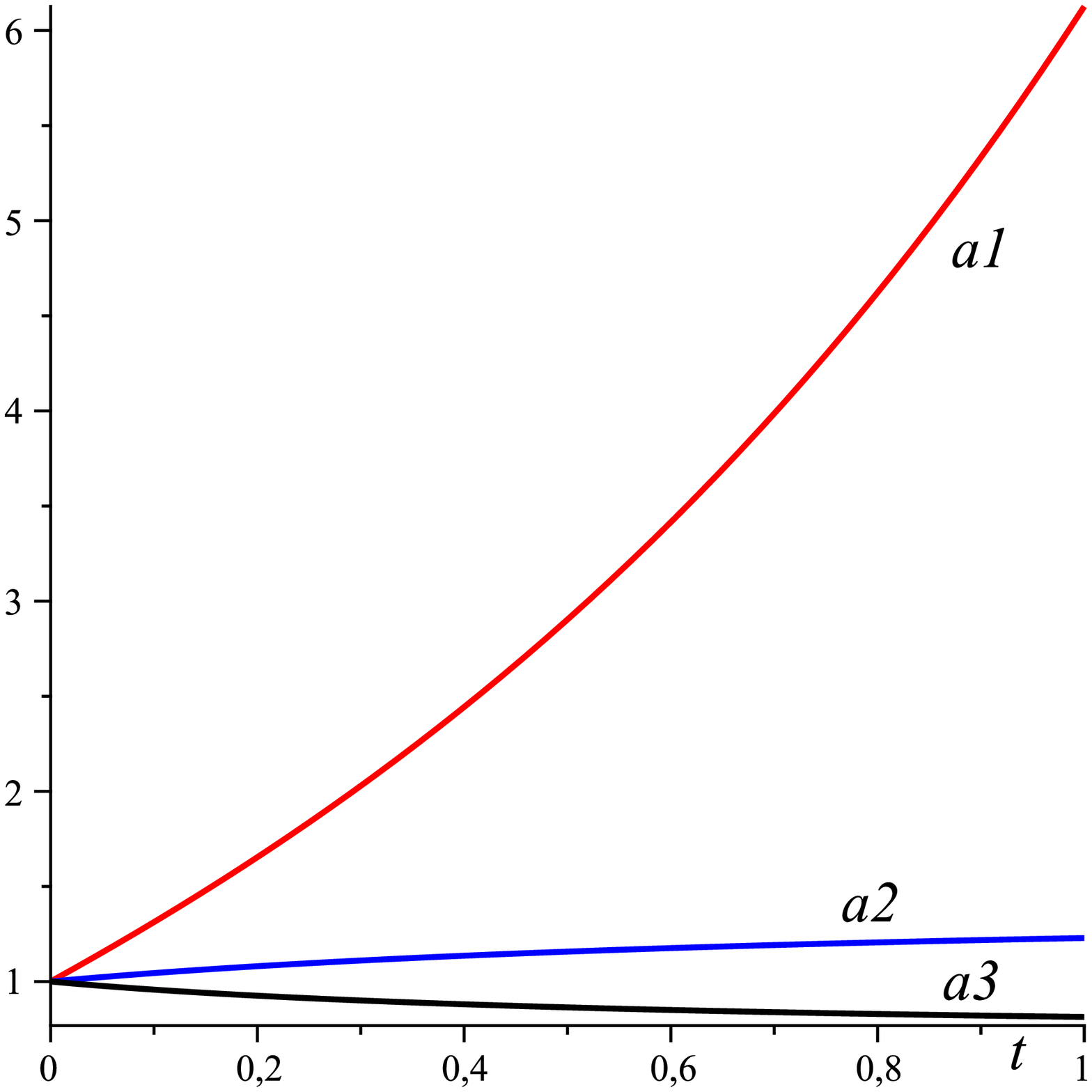}{0.45}{Evolution of metric functions for $\alpha
= 1.5$.} {0.45}{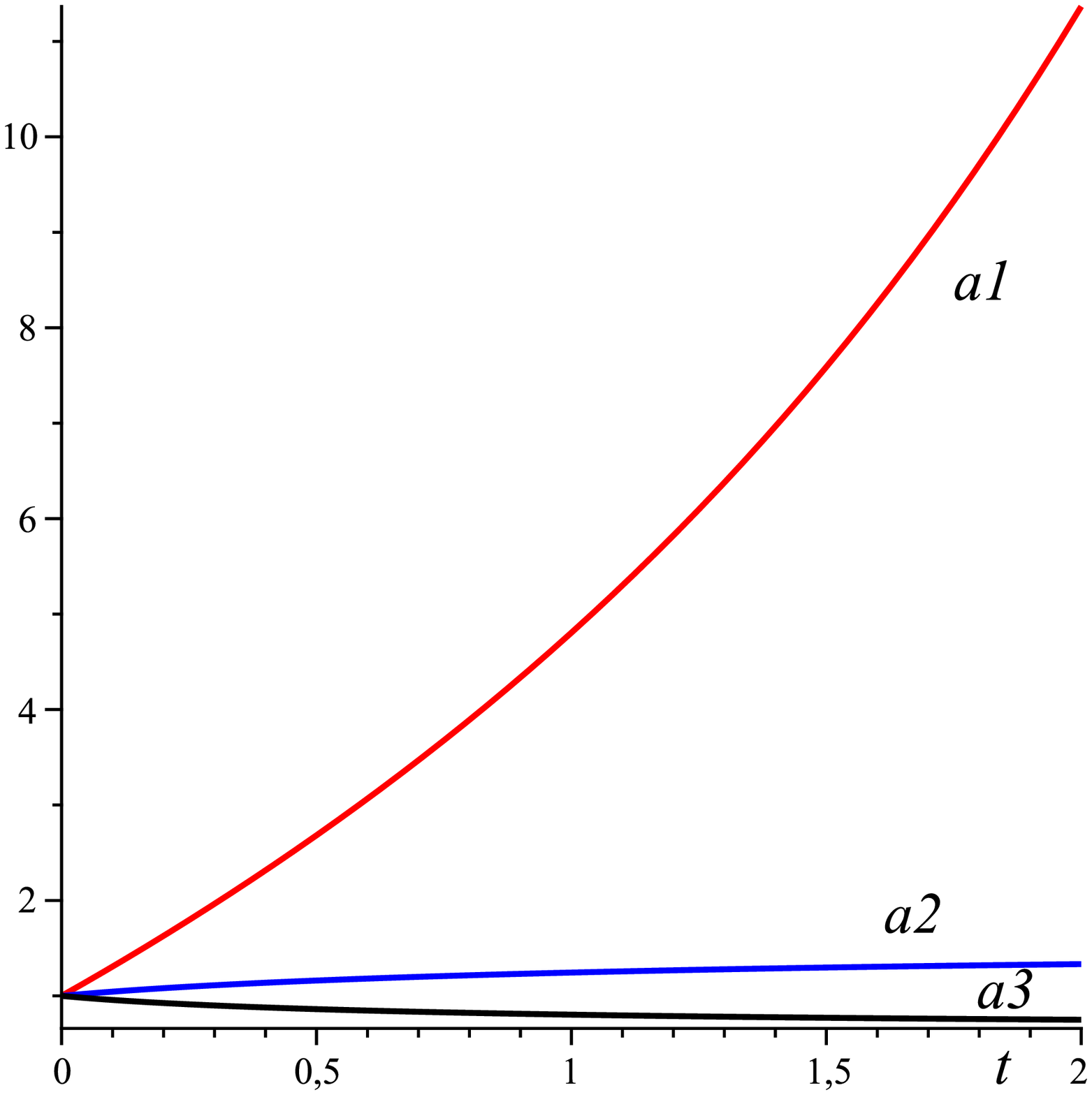}{0.45}{Evolution of metric functions for
$\alpha =  -1$.}{0.45}

In Figs. \ref{abc+105} and \ref{abc-1} we plot the behavior of
metric functions for positive and negative values of $\alpha$. As is
seen from the figures, with the expansion of the Universe $a_1$
increases exponentially, while $a_2, a_3$ tend to constants. Thus we
see that introduction of cosmic string does not allow isotropization
of initially anisotropic space-time. In the following subsections we
discuss the singularity problem and isotropization process in
detail.

\subsection{Singularities}

The next task is to investigate the space-time singularities in the
present model. Let us rewrite the metric functions in the following
form:
\begin{equation}
a_i = C_i \tau^{N_i} \exp \biggl(Y_i \int
\frac{dt'}{\tau}\biggr)\,,\label{ai}
\end{equation}
where $C_1 = A_1$,\, $C_2 = \sqrt{D/A_1}$,\, $C_3 =
1/\sqrt{DA_1}$,\, $N_1 = +1$,\,$N_2 = N_3 = 0$,\,$Y_1 = 0$,\,
$Y_2 = X/2$, and $Y_3 = -X/2$. Then the first and the second
derivatives of the metric functions take the form:
\begin{subequations}
\label{singvis}
\begin{eqnarray}
\frac{\dot a_i}{a_i} &=& N_i \frac{\dot{\tau}}{\tau} +
\frac{Y_i}{\tau}\,, \label{singvis1}\\
\frac{\ddot a_i}{a_i} &=& N_i \frac{\ddot{\tau}}{\tau} + (N_i^2 -
N_i) \Bigl(\frac{\dot{\tau}}{\tau}\Bigr)^2 + (2N_i - 1) Y_i \frac{
\dot{\tau}}{\tau^2} + \frac{Y_i^2}{\tau^2}\,.
\end{eqnarray}
\end{subequations}

We study the singularities analyzing the regularity properties of
the  Krets\-ch\-mann scalar which for the metric \eqref{BI} reads
\begin{equation}
\mathcal{K} = 4\Bigl[\Bigl(\frac{\ddot a_1}{a_1}\Bigr)^2 + \Bigl(\frac{\ddot
a_2}{a_2}\Bigr)^2 + \Bigl(\frac{\ddot a_3}{a_3}\Bigr)^2 +
\Bigl(\frac{\dot a_1}{a_1}\frac{\dot a_2}{a_2}\Bigr)^2 +
\Bigl(\frac{\dot a_2}{a_2}\frac{\dot a_3}{a_3}\Bigr)^2 +
\Bigl(\frac{\dot a_3}{a_3}\frac{\dot a_1}{a_1}\Bigr)^2\Bigr]\,.
\label{Kretsch}
\end{equation}
Evidently, a singularity can occur when some or all scale factors $a_i$
tend to zero or infinity. We shall follow the criteria given in
\cite{Bronshik} which states:

\bigskip

(i) $t$ finite,  some $a_i \to \infty$. {\it If at least one scale
factor becomes infinite at finite $t$, it is a curvature
singularity.}

(ii) $t$ finite, some $a_i \to 0$. {\it If more than one scale
factor turns to zero at finite $t$, it is a singularity. If only one
scale factor is zero at finite $t$, the spacetime can be
nonsingular.}

(iii) $t \to \infty$, some $a_i \to \infty$. {\it Such an asymptotic
can only be singular if at least one scale factor grows faster than
exponentially:}\\
$a_i(t) \gg \exp{(k|t|)},\quad k = {\rm const.} > 0.$

(iv)$t \to \infty$, some $a_i \to 0$. {\it Such an asymptotic can
only be singular if at least one scale factor vanishes faster than
exponentially:} \\
$a_i(t) \ll \mathcal{O}(\exp{[-k|t|]}),\quad k = {\rm const.} > 0.$

\bigskip

Looking at the asymptotic behaviors described above (Section
\ref{asymptotic}) we conclude that in the spinor field approach of the
cosmic strings the present model is nonsingular at the end of the
evolution. No scale factor of the BI metric presents a growth faster
than exponentially or vanishing faster than exponentially for
$t \to \infty$.

\subsection{Isotropization}

Since the present-day Universe is surprisingly isotropic, it is
important to see whether our anisotropic BI model evolves into an
isotropic FRW model. Isotropization means that at large physical
times $t$, when the volume factor $\tau$ tends to infinity, the
three scale factors $a_i(t)$ grow at the same rate. Two wide-spread
criteria of isotropization read
\begin{subequations}
\label{aniso}
\begin{eqnarray}
{\cal A} &=&  \frac{1}{3} \sum\limits_{i=1}^{3} \frac{H_i^2}{H^2} -
1  \to 0\,,\\
\Sigma^2 &=& \frac{1}{2}  {\cal A} H^2 \to 0\,.
\end{eqnarray}
\end{subequations}
Here ${\cal A}$ and $\Sigma^2$ are the average anisotropy and
shear, respectively, while $H_i = \dot{a_i}/a_i$ is the directional
Hubble parameter evaluated in \eqref{singvis1}. We also investigate
the isotropization condition proposed in \cite{Bronshik}
\begin{equation}
\frac{a_i}{a}\Bigl|_{t \to \infty} \to {\rm const}\,, \label{isocon}
\end{equation}
where $a(t) = \tau^{1/3}$ is the average scale factor.
Provided that condition \eqref{isocon} is fulfilled, by rescaling
some of the coordinates, we can make $a_i/a \to 1$, and the metric
will become manifestly isotropic at large $t$.

In order to study the isotropization process in the present
model, we evaluate from \eqref{ai}
\begin{equation}
\frac{a_i}{a} = \frac{a_i}{\tau^{1/3}} =  C_i \tau^{N_i-1/3} \exp
\biggl(Y_i \int \frac{dt'}{\tau}\biggr)\,,\label{isocon1}
\end{equation}
Taking into account the value of $N_i$ and $Y_i$ we see that $a_i/a$
do not tend to constants as $t \to \infty$.
For example, in the generic case with the exponential expansion of the
volume of the Universe \eqref{tauexp},
$a_1/a \to \infty$,  $a_2/a \to 0$ and $a_3/a \to 0$.
So in the case considered, no
isotropization process takes place. Indeed, at the early stage of
evolution, where $\tau \to 0$ we assume $a_1 \to \infty$,\, $a_2 \to
0$ and $a_3 \to 0$, that is at this stage the Universe looks like a
one-dimensional string. In the asymptotic region where $t \to
\infty$ and $\tau \to \infty$ we have $a_1 \to \infty$,  while
$a_2$ and $a_3$ evaluating to finite values.

Let us also examine the possibility of an isotropization process
using conditions \eqref{aniso}. In view of \eqref{singvis} from
\eqref{aniso} in the generic case \eqref{tauexp} one finds:
\begin{subequations}
\label{aniso1}
\begin{eqnarray}
{\cal A} &=&  2 \,,\\
\Sigma^2 &=& \frac{1}{9} \,.
\end{eqnarray}
\end{subequations}

As one sees, both ${\cal A}$ and $\Sigma^2$ are some
positive defined, non vanishing quantities.
Thus, in general,  the present model does not undergo
an isotropization process, the anisotropic feature of the Universe
from the early stages being retained.

\section{Conclusions}

We have studied the evolution of a anisotropic universe given by a
BI cosmological model in presence of  cosmic strings and
magnetic field. In doing so we exploit the spinor simulation of
cosmic strings that allows us to solve the system of Einstein's
equations. Exact solutions have been supplemented with some
numerical evaluations. It is shown that the presence of cosmic
strings does not allow the anisotropic Universe to evolve into an
isotropic one.

\subsection*{Acknowledgments}
The authors gratefully acknowledge the support from the
joint Romanian-LIT, JINR, Dubna Research
Project, theme no. 05-6-1060-2005/2010. M.V. is partially supported by
the CNCSIS Program IDEI - 571/2008, Romania.

%

\begin{thebibliography}{99}
%
\bibitem{misner} C.W. Misner, {\it The Astrophys. J.} {\bf 151}, 431 (1968).
%
\bibitem{guth} A. Guth, {\it Phys. Rev. D} {\bf 23}, 347 (1981).
%
\bibitem{henprd} M. Henneaux, {\it Phys. Rev. D}  {\bf 21}, 857 (1980).
%
\bibitem{sahaprd} B. Saha, {\it Phys. Rev. D} {\bf 64}, 123501 (2001).
%
\bibitem{greene} C. Armend\'{a}riz-Pic\'{o}n and P.B. Greene,
{\it General Relat. Grav.} {\bf 35}, 1637 (2003).
%
\bibitem{SBprd04} B. Saha and T. Boyadjiev, {\it Phys. Rev. D} {\bf 69},
124010 (2004).
%
\bibitem{kremer1} V.O. Ribas, F.P. Devecchi, and G.M. Kremer,
{\it Phys. Rev. D} {\bf 72}, 123502 (2005).
%
\bibitem{ECAA06}  B. Saha, {\it Physics of Particles and
Nuclei} {\bf 37}. Suppl. 1, S13 (2006).
%
\bibitem{sahaprd06} B. Saha, {\it Phys. Rev. D} {\bf 74}, 124030 (2006).
%
\bibitem{BVI} B. Saha, {\it Phys. Rev. D} {\bf 69}, 124006 (2004).
%
\bibitem{shikin} V.G. Krechet, M.L. Fel'chenkov and G.N. Shikin,
{\it Grav. Cosmology} {\bf 14}, 292 (2008).
%
\bibitem{spinpf0}
B. Saha, {\it Spinor model of a perfect fluid }, arXiv: 0901.1387
[gr-qc].
%
\bibitem{spinpf}
B. Saha, {\it Spinor model of a perfect fluid: examples}, arXiv:
0902.2097 [gr-qc].

\bibitem{KM} T. Koivisto and D.F. Mota, {\it Phys. Lett. B} {\bf 644},
104 (2007).
%
\bibitem{KM1} T. Koivisto and D.F. Mota, {\it Phys. Rev. D} {\bf 75},
023518 (2007).
%
\bibitem{letelier}
P.S. Letelier, {\it Phys. Rev. D} {\bf 28}, 2414 (1983).
%
\bibitem{pradhan}
A. Pradhan, A. K. Yadav, R. P. Singh and V. K. Singh,
{\it Astrophys. Space Sci.} {\bf 312}, 145 (2007).
%
\bibitem{tade}
G.S. Khadekar and S.D. Tade,
{\it Astrophys. Space Sci.} {\bf 310}, 47 (2007).
%
\bibitem{lich} A. Lichnerowicz,
{\it Relativistic Hydrodynamics and Magnetohydrodynamics},
(Benjamin, New York, 1967).
%
\bibitem{ass}
B. Saha, {\it Astrophys. Space Sci.} {\bf 299}, 149 (2005).
%
\bibitem{SV}
B. Saha and M. Visinescu, {\it Astrophys. Space Sci.} {\bf
315}, 99 (2008).
\bibitem{SRV}
B. Saha, V. Rikhvitsky and M. Visinescu, {\it Cent. Eur. J. Phys.}
{\bf 8}, 113 (2010).
%
\bibitem{Bronshik} K.A. Bronnikov, E.N. Chudaeva and G.N. Shikin,
Class. Quantum Grav. {\bf 21}, 3389 (2004).
\end{thebibliography}
\end{document}